\documentclass[a4paper,12pt]{article}
\usepackage{fullpage}
\usepackage{amsmath, amsthm, amssymb, mathrsfs, graphicx}
\usepackage{ifthen, url, subfigure, epic, amsfonts}
\usepackage[usenames]{color}

\usepackage[sort&compress]{natbib}
\bibpunct{(}{)}{;}{a}{}{,}

\theoremstyle{plain}

\theoremstyle{remark}

\theoremstyle{definition}

\theoremstyle{remark}
\newtheorem{ex}{Example}

\theoremstyle{remark}

\theoremstyle{definition}
\newtheorem*{PRalgorithm}{Predictive recursion algorithm}

\theoremstyle{definition}
\newtheorem*{SAalgorithm}{Simulated annealing algorithm}

\newcommand{\U}{\mathscr{U}}
\newcommand{\Ubar}{\overline{\mathscr{U}}}
\newcommand{\Y}{\mathscr{Y}}

\newcommand{\FF}{\mathbb{F}}
\newcommand{\FFbar}{\overline{\mathbb{F}}}
\newcommand{\MM}{\mathbb{M}}
\newcommand{\RR}{\mathbb{R}}

\newcommand{\E}{\mathsf{E}}

\newcommand{\prob}{\mathsf{P}}

\renewcommand{\phi}{\varphi}

\newcommand{\iid}{\overset{\text{\tiny iid}}{\sim}}

\begin{document}

\title{An approximate Bayesian marginal likelihood approach for estimating finite mixtures}
\author{
Ryan Martin \\  
Department of Mathematics, Statistics, and Computer Science \\ 
University of Illinois at Chicago \\ 
\url{rgmartin@math.uic.edu} 
}
\date{\today}

\maketitle

\begin{abstract} 
Estimation of finite mixture models when the mixing distribution support is unknown is an important problem.  This paper gives a new approach based on a marginal likelihood for the unknown support.  Motivated by a Bayesian Dirichlet prior model, a computationally efficient stochastic approximation version of the marginal likelihood is proposed and large-sample theory is presented.  By restricting the support to a finite grid, a simulated annealing method is employed to maximize the marginal likelihood and estimate the support.  Real and simulated data examples show that this novel stochastic approximation--simulated annealing procedure compares favorably to existing methods.  

\medskip

\end{abstract}

\section{Introduction}
\label{S:intro}

It is well-known that complicated data sets can be described by a mixture of a few relatively simple models.  The catch, however, is that this finite mixing distribution is generally difficult to specify.  For example, in clustering or empirical Bayes problems, the mixing distribution is exactly the quantity of interest.  Therefore, estimating the unknown mixing distribution is an important problem.  Key references include \citet{titterington}, \citet{mclachlanbasford}, and \citet{mclachlanpeel}.  

To fix notation, we assume that data $Y_1,\ldots,Y_n$ are independent observations from a common distribution with density $m(y)$, modeled as a finite mixture 
\begin{equation}
\label{eq:finite.mixture}
m_{f,U}(y) = \sum_{u \in U} p(y \mid u) f(u), \quad y \in \Y, \quad U \subset \Ubar, 
\end{equation}
where $\Ubar$ is a known compact set and $(y,u) \mapsto p(y \mid u)$ is a known kernel on $\Y \times \Ubar$, such as Gaussian or Poisson.  The goal is to estimate the unknown finite support set $U$ and the corresponding mixing weights $f=\{f(u): u \in U\}$.  

A classical approach to this problem is nonparametric maximum likelihood \citep{lindsay1995}, where the unknowns $(f,U)$ are estimated by maximizing the likelihood function $\prod_{i=1}^n m_{f,U}(Y_i)$.  The goal of maximum likelihood is simply to produce an estimate of $m_{f,U}$ that fits the data well, so there are no built-in concerns about the size of the estimated support.  But as \citet{priebe} argues, there are compelling reasons to seek a well-fitting mixture with as few support points as possible.  Since these two considerations---good fit and small support---are at odds with one another, a modification of maximum likelihood/minimum distance estimation is needed.  The typical non-Bayesian strategy is to penalize the likelihood function of those models whose mixing distribution support is too large.  This choice of penalty can be simple, like the AIC \citep{akaike1973} or BIC \citep{schwarz1978} penalties discussed by \citet{leroux}, or can be more sophisticated, like the SCAD penalty \citep{fanli2001} used in the mixture model context by \citet{chen.khalili.2008}.  Minimum distance methods with similar support size penalties have also been proposed, e.g., by \citet{james} and \citet{woosriram2006, woosriram2007}.  In the Bayesian context, there are a number of other methods.  \citet{ishwaran.james.sun.2001} give a solution based on Bayes factors, \citet{richardson} presents a method in which each model is embedded into a larger parameter space on which a prior distribution is imposed, and \citet{roederwasserman} describe a practical solution based on the posterior distribution of the number of mixture components.  A typical nonparametric Bayesian strategy is to model the mixing distribution itself as a random draw from a Dirichlet process distribution \citep{ferguson1973}.  Discreteness properties of the Dirichlet process imply that the distribution of the observables is almost surely a finite mixture, where the number of mixture components, as well as the component-specific parameters, are random quantities.  This flexible modeling strategy effectively allows the data to determine the mixture structure.  Efficient Markov chain Monte Carlo algorithms \citep{escobar.west.1995, maceachern.muller.1998, neal2000} are now available for fitting the Dirichlet process mixture model to data; see, e.g., \citet{muller.quintana.2004}.  
 
In this paper I introduce a new approach for fitting finite mixture models \eqref{eq:finite.mixture} with unknown support size.  The starting point is the construction of a marginal likelihood for the mixing distribution support based on a Bayesian hierarchical model.  For any fixed support set $U$, a computationally efficient approximation of the Bayesian marginal likelihood is available, based on a Robbins--Monro type of stochastic approximation algorithm called \emph{predictive recursion}.  But despite its efficiency, estimating $U$ by maximizing this approximate marginal likelihood over \emph{all possible} $U$ is not feasible.  The key idea is to chop up the bounding set $\Ubar$ into a large, suitably fine grid $\U$, and search for the best approximation to the underlying density $m$ by mixtures supported on subsets of $\U$.  Thus, the essentially nonparametric problem is transformed to a high-dimensional parametric one.  The parameter $U$ takes values in a very large finite set and for this high-dimensional combinatorial optimization problem, we propose a fast simulated annealing procedure.  This novel combination of stochastic approximation and simulated annealing for finite mixture model estimation shall be called SASA.     

Asymptotic convergence properties of the SASA estimates are presented in Section~\ref{SS:theory}.  Specifically, for given $\U$, the SASA procedure will asymptotically identify the best mixture over all those supported on subsets of $\U$.  Here ``best'' is measured in terms of the Kullback--Leibler divergence, so that SASA acts asymptotically like a minimum distance estimation method.  SASA also achieves the optimal rate of convergence obtained in \citet{chen1995}, although I suspect that the restriction to finitely many $U$'s actually yields faster rates.  Furthermore, unlike the estimates of \citet{james} or \citet{woosriram2006}, the SASA estimate of the support size will always converge to a finite number, and will be consistent if the true mixing distribution support is a subset of $\U$.  

For flexibility, the finite set $\U$ of candidate support points should be large.  But it is often the case that one believes that the true support size is considerably smaller than $|\U|$.  To account for these prior beliefs, I recommend a regularized version of the approximate marginal likelihood that penalizes supports $U$ in $\U$ which are too large.  In particular, I suggest a penalty determined by a binomial prior on $|U|$, with success probability parameter chosen to reflect the user's belief about the true mixture complexity.

The SASA approach can, in principle, handle mixtures over any number of parameters, but computation can be relatively expensive for mixtures over two or more parameters.  In Section~\ref{S:locscale}, I modify the proposed algorithm to give a fast approximation to the SASA solution in finite location-scale mixtures.  This approximation focuses on a justifiable class of admissible subsets and this restriction can substantially decrease the complexity of the combinatorial optimization problem to solve.

\section{Likelihood for the mixing distribution support}
\label{S:marg.lik}

The following model selection reasoning is ubiquitous in statistics: calculate a ``score'' for each model in consideration and then pick the model with the best score.  For example, the model score is often the maximized likelihood function over the model's parameters.  But the case can be made that this maximized (profile) likelihood for the model essentially ignores the uncertainty in estimating the model parameters.  In such cases, a marginal likelihood, with model parameters ``integrated out'' may be more reasonable, justifying a sort of Bayesian perspective.  Along these lines, take the mixing distribution support set $U$ to be fixed for the time being and consider the following hierarchical model:
\begin{equation}
\label{eq:hierarchy}
Y_1,\ldots,Y_n \mid (f, U) \iid m_{f,U}, \quad \text{and} \quad f \mid U \sim \prob_U,
\end{equation}
where $m_{f,U}$ is as in \eqref{eq:finite.mixture}, and $\prob_U$ is a generic prior for the random discrete distribution $f$ supported on $U$.  While other choices are possible, I will henceforth assume that $\prob_U$ is a finite-dimensional Dirichlet distribution on $U$ with precision parameter $\alpha_0 > 0$ and base measure $f_{0,U}$, a probability vector indexed by $U$.  For this model, the marginal likelihood for $U$ is of the form 
\begin{equation}
\label{eq:dp-marg}
L_n^\dagger(U) = \int \Bigl\{ \prod_{i=1}^n m_{f,U}(Y_i) \Bigr\} \,d\prob_U(f) = \prod_{i=1}^n \sum_{u \in U} p(Y_i \mid u) \hat{f}_{i-1,U}(u), 
\end{equation}
where $\hat{f}_{i-1,U} = \E_U(f \mid Y_1,\ldots,Y_{i-1})$ is the posterior mean.  The second equality in \eqref{eq:dp-marg} is a consequence of Fubini's theorem and the linearity of the mixture.  This likelihood can be computed efficiently, without explicitly evaluating the $\hat f_{i-1,U}$'s, via the sequential imputation method of \citet{jsliu1996}.  The function $L_n^\dagger(U)$ is a genuine likelihood for $U$ in the sense that it defines a reasonable, data-dependent ranking of candidate support sets, properly accounting for the uncertainty in the mixing distribution weights $f$.  So $L_n^\dagger$ can be used to assign a preference between two supports $U$ and $U'$, but I claim that it can also be used, in a natural way, to estimate the support, up to an approximation.  

Suppose, first, that the unknown support set $U$ is known to be contained in a compact set $\Ubar$.  This is a very weak assumption that can always be justified in practice.  Now chop up $\Ubar$ into a sufficiently fine finite grid $\U$ such that assuming $U \subseteq \U$ is no practical restriction, in the sense that the data-generating density $m(y)$ can be closely approximated by some mixture supported on a subset of $\U$.  In the examples that follow, good solutions can be found when the grid $\U$ is of only moderate size.  The advantage of this approximation is that the essentially nonparametric problem of estimating the unknown support becomes a very-high-dimensional parametric problem.  In fact, there are only finitely many possible parameter values so theoretical convergence of estimators follows from simple point-wise convergence of (a normalized version of) the marginal likelihood.  The drawback, however, is that maximizing the marginal likelihood over $\U$ is a relatively challenging combinatorial optimization problem.  Although \eqref{eq:hierarchy} is a reasonable model, it turns out that this fully Bayesian framework is not completely satisfactory from a computational point of view; see Example~\ref{ex:galaxy0} below.  Next I propose a second approximation which closely mimics the Bayesian results at a fraction of the computational cost.

\section{The SASA method for finite mixtures}
\label{S:prmlfinite} 

\subsection{A stochastic approximation-based likelihood}
\label{SS:unknown}

First consider the general problem where the common marginal density $m(y)$ for $\Y$-valued observations $Y_1,\ldots,Y_n$ is modeled as a nonparametric mixture 
\begin{equation}
\label{eq:np.mixture}
m_f(y) = \int_{\U} p(y \mid u) f(u) \,d\nu(u), \quad y \in \Y, 
\end{equation}
where $\U$ is a known set, not necessarily finite, and $f \in \FF$ is the unknown mixing density to be estimated.  Here $\FF = \FF(\U,\nu)$ is the set of all densities with respect to a $\sigma$-finite Borel measure $\nu$ on $\U$.  \citet{newton02} presents the following \emph{predictive recursion} algorithm for nonparametric estimation of $f$.  

\begin{PRalgorithm}
Fix $f_0 \in \FF$ and a sequence of weights $\{w_i: i \geq 1\} \subset (0,1)$.  For $i=1,\ldots,n$, compute $m_{i-1}(y) = m_{f_{i-1}}(y)$ as in \eqref{eq:np.mixture} and 
\begin{equation}
\label{eq:recursion}
f_i(u) = (1-w_i) f_{i-1}(u) + w_i \, p(Y_i \mid u) f_{i-1}(u) \, / \, m_{i-1}(Y_i). 
\end{equation}
Then return $f_n$ and $m_n$ as the final estimates of $f$ and $m$, respectively.  
\end{PRalgorithm}

\citet{martinghosh} showed that $\{f_n\}$ is a Robbins--Monro stochastic approximation process.  Key properties of predictive recursion include its fast computation and its ability to estimate a mixing density $f$ absolutely continuous with respect to any user-defined dominating measure $\nu$.  That is, unlike the nonparametric maximum likelihood estimate which is almost surely discrete \citep[][Theorem~21]{lindsay1995}, $f_n$ can be discrete, continuous, or both, depending on $\nu$.  Herein I shall take $\nu$ to be counting measure on a finite set $\U$, but see \citet{mt-prml, mt-test} for applications of predictive recursion where $\nu$ is continuous or both discrete and continuous.  

Large-sample properties of $f_n$ and $m_n$ can be obtained under fairly mild conditions on the kernel $p(y \mid u)$ and the true data-generating density $m(y)$.  Let $\MM$ denote the set of all mixtures $m_f$ in \eqref{eq:np.mixture} as $f$ ranges over $\FF$, and for two densities $m$ and $m'$ let $K(m,m') = \int \log\{m(y)/m'(y)\} m(y)\,dy$ denote the Kullback--Leibler divergence of $m'$ from $m$.  Then \citet{tmg} prove almost sure $L_1$ and weak convergence of $m_n$ and $f_n$, respectively, when $m \in \MM$.  When $m \not\in \MM$, \citet{mt-rate} show that there exists a mixing density $f^\star$ in $\FFbar$, the weak closure of $\FF$, such that $K(m,m_{f^\star}) = \inf\{K(m,m_f): f \in \FFbar\}$, and $m_n$ converges almost surely in $L_1$ to $m_{f^\star}$.  As a corollary, they show that if the mixture \eqref{eq:np.mixture} is identifiable, then $f_n$ converges weakly to $f^\star$ almost surely.  Moreover, for a certain choice of weights $\{w_n\}$, they obtain a conservative $o(n^{-1/6})$ bound on the rate at which $m_n$ converges to $m_{f^\star}$.  The rate for $f_n$ in the general case is unknown, but \citet{pr-finite} obtains a near parametric $n^{-1/2}$ rate for $f_n$ in the finite mixture case; see also Section~\ref{SS:theory} below.   

Define a stochastic approximation-based marginal likelihood
\begin{equation}
\label{eq:prml}
L_n(U) = \prod_{i=1}^n m_{i-1,U}(Y_i) = \prod_{i=1}^n \sum_{u \in U} p(Y_i \mid u) f_{i-1}(u). 
\end{equation}
This is based on an interpretation of $m_{i-1,U}(Y_i)$ as the conditional density of $Y_i$ given $Y_1,\ldots,Y_{i-1}$.  I claim that $L_n(U)$ is an approximation of Dirichlet prior Bayes marginal likelihood $L_n^\dagger(U)$.  Towards this, recall that $\hat f_{k,U} = \E_U(f \mid Y_1,\ldots,Y_k)$ is the posterior mean of the mixing distribution on fixed $U$, given $Y_1,\ldots,Y_k$.  Then $\sum_{u \in U} p(Y_i \mid u) \hat f_{i-1,U}(u)$ is the conditional density of $Y_i$ given $Y_1,\ldots,Y_{i-1}$ based on the Dirichlet hierarchical model.  Also the Polya urn representation of the Dirichlet distribution \citep[][Sec.~3.1.2]{ghoshramamoorthi} implies that
\[ \hat f_{1,U}(u) = \frac{\alpha_0}{\alpha_0 + 1} f_{0,U}(u) + \frac{1}{\alpha_0 + 1} \, \frac{p(Y_1 \mid u) f_{0,U}(u)}{\sum_{u' \in U} p(Y_1 \mid u') f_{0,U}(u')}, \]
a mixture of the prior guess and a predictive distribution on $U$ given $Y_1$.  If $\alpha_0 = 1/w_1-1$, then $\hat f_{1,U}(u)$ is exactly $f_1(u)$ in \eqref{eq:recursion}.  This correspondence holds exactly only for a single observation, but \citet{mt-prml} argue that $f_{i-1,U}$, for any $i$, acts as a dynamic, mean-preserving filter approximation to the Bayes estimate $\hat f_{i-1,U}$.  Then $L_n(U)$ in \eqref{eq:prml} can be viewed as a plug-in approximation of the Bayes marginal likelihood $L_n^\dagger(U)$ in \eqref{eq:dp-marg}, with $f_{i-1,U}$ in place of the Bayes estimate $\hat f_{i-1,U}$.  See, also, Example~\ref{ex:galaxy0} below.

In what follows, I will work with the marginal likelihood on the log-scale,
\begin{equation}
\label{eq:lprml}
\ell_n(U) = \log L_n(U) = \sum_{i=1}^n \log \Bigl\{ \sum_{u \in U} p(Y_i \mid u) f_{i-1}(u) \Bigr\}.
\end{equation}
The goal is to estimate $U$ by maximizing $\ell_n(U)$ over $U$.  Restricting $U$ to be a subset of the finite set $\U$ is a helpful first step, but even when $|\U|$ is only moderately large, the set of possible supports is still enormous, cardinality $2^{|\U|}-1$.  So despite the fact that the search space is finite, its size makes this is challenging problem.  In Section~\ref{SS:sann}, I give a simulated annealing algorithm to solve this combinatorial optimization problem.

\subsection{Optimization with simulated annealing}
\label{SS:sann}

As described above, maximizing $\ell_n(U)$ over all subsets $U \in 2^{\U}$ is a combinatorial optimization problem.  The challenge is that $2^{\U}$ is so large that it is not feasible to evaluate $\ell_n(U)$ for each $U$.  Simulated annealing is a stochastic algorithm where, at iteration $t$, a move from the current state $U^{(t)}$ to a new state $U^{(t+1)}$ is proposed so that $\ell_n(U^{(t+1)})$ will tend to be larger than $\ell_n(U^{(t)})$.  An important feature of simulated annealing is the decreasing temperature sequence $\{\tau_t: t \geq 0\}$.  Following \citet{hajek1988} and \citet{belisle1992}, I take the default choice $\tau_t = a/\log(1+t)$ for a suitable $a$, chosen by trial-and-error.  For the numerical examples that follow, $a=1$ gives acceptable results.  

To simplify the discussion, to each subset $U \subset \U = \{u_1,\ldots,u_S\}$, where $S = |\U|$, associate a binary $S$-vector $H \in \{0,1\}^S$.  Then $H_s = 1$ if $u_s \in U$ and $H_s = 0$ otherwise.  In other words, $H_s$ determines whether $u_s$ is in or out of the mixture.  It clearly suffices to define the optimization of $\ell_n(U)$ over $2^{\U}$ in terms of the $H$ vectors.   Then the simulated annealing algorithm goes as follows.   

\begin{SAalgorithm}
Choose a starting point $H^{(0)}$ and a maximum number of iterations $T$.  Set $t=1$ and generate a sequence $\{H^{(t)}: t=1,\ldots,T\}$ as follows:  
\begin{enumerate}
\item Simulate $H_{\text{new}}$ from a probability distribution $\pi^{(t)}$ on $\{0,1\}^S$, possibly depending on $t$ and the current iterate $H^{(t)}$.  
\vspace{-2mm}
\item Define the acceptance probability 
\[ \alpha(t) = 1 \wedge \exp \bigl[ \bigl\{ \ell_n(H_{\text{new}}) - \ell_n(H^{(t)}) \bigr\} / \tau_t \bigl],  \]
where $\ell_n(H)$ is the PR marginal likelihood defined in \eqref{eq:prml}, written as a function of the indicator $H$ that characterizes $U$, and set
\[ H^{(t+1)} = \begin{cases} H_{\text{new}} & \text{with probability $\alpha(t)$} \\ H^{(t)} & \text{with probability $1-\alpha(t)$} \end{cases} \]
\item If $t < T$, set $t \leftarrow t+1$ and return to Step~1; else, exit the loop.  
\end{enumerate}
Then return the visited $H^{(t)}$ with the largest log-likelihood $\ell_n(H^{(t)})$. 
\end{SAalgorithm}

Herein, the initial choice is $H_s^{(0)} = 1$ for each $s$, which corresponds to the full mixture.  The key to the success of simulated annealing is that while all uphill moves are taken, some downhill moves, to ``less likely'' $U_{\text{new}}$, are allowed through the flip of a $\alpha(t)$-coin in Step~2.  This helps prevent the algorithm from getting stuck at local modes.  But the vanishing temperature $\tau_t$ makes these downhill moves less likely when $t$ is large.  

It remains to specify a proposal distribution $\pi^{(t)}$ in Step~1.  I shall assume that a draw $H_{\text{new}}$ from $\pi^{(t)}$ differs from $H^{(t)}$ in exactly $k \geq 1$ positions.  In other words, $k$ of the $S$ components of $H^{(t)}$ are chosen and then each is flipped from 0 to 1 or from 1 to 0.  The choice of components is not made uniformly, however.  To encourage a solution with a relatively small number of mixture components, I want $\pi^{(t)}$ to assign greater mass to those components $H_s^{(t)}$ in $H^{(t)}$ such that $H_s^{(t)} = 1$.  The particular choice of weights is 
\begin{equation}
\label{eq:proposal0}
\pi_s^{(t)} \propto 1+ \Bigl( \frac{S}{\sum_{s=1}^S H_s^{(t)}} \Bigr)^r \cdot H_s^{(t)}, \quad s=1,\ldots,S, \quad r \geq 1. 
\end{equation}
Note that when most of the components of $H^{(t)}$ are 1, equivalently, when $|U^{(t)}|$ is large, the sampling is near uniform, whereas, when $H^{(t)}$ is sparse, those components with value 1 have a greater chance of being selected.  

Next I discuss a few miscellaneous computational details.  
\begin{itemize}
\item The log marginal likelihood $\ell_n(U)$ depends on the order in which the data $Y_1,\ldots,Y_n$ are processed.  To reduce this dependence, I take $\ell_n(U)$ to be the average of the log marginal likelihoods over several random data permutations.  The speed of predictive recursion for fixed $U$ makes this permutation-averaging computationally feasible.  Herein I use 100 permutations but, in my experience, the SASA estimates are relatively stable for as few as 25 permutations.  These permutations are chosen once and kept fixed throughout optimization process.  
\vspace{-2mm}
\item To avoid various degeneracies, specifically in \eqref{eq:proposal0}, I set $\ell_n(\varnothing) = -\infty$.  This means that if a move to an empty mixing distribution support is proposed, then it will surely be rejected since the corresponding $\alpha(t)$ would be zero.  
\vspace{-2mm}
\item For all examples, the ``distance'' $k$ between two consecutive support sets, is taken to be 1.  Also, choosing $r=1$ in \eqref{eq:proposal0} works well.  
\vspace{-2mm}
\item For all examples, I run simulated annealing for $T=5000$ iterations.  As with the choice of permutations, my choice here is rather conservative, as the estimates are often relatively stable for $T=2000$.  
\end{itemize}  
R codes to implement the SASA procedure are available at my website \url{www.math.uic.edu/~rgmartin/research.html}.  The R function {\tt optim}, with the option {\tt method="SANN"}, is the driving force behind the simulated annealing.  A C subroutine used to efficiently evaluate the log marginal likelihood $\ell_n(U)$ for any fixed $U$.

\subsection{Large-sample theory}
\label{SS:theory}

Suppose that $\U$ is a fixed finite set and $U \subseteq \U$ is any subset.  Assume that the predictive recursion weights $\{w_i: i \geq 1\}$ are given by $w_i = (i+1)^{-\gamma}$ for some $\gamma \in (0.5,1)$.  With $U$ fixed, convergence of $f_{n,U}$ at a $n^{-(1-1/2\gamma)}$ rate is established in \citet{pr-finite}.  This result holds only for fixed $U$, while the present focus is on the case of unknown $U$.  Towards this, consider a normalized version of the log marginal likelihood $\ell_n(U)$, namely 
\[ K_n(U) = \frac1n \sum_{i=1}^n \log \frac{m(Y_i)}{m_{i-1,U}(Y_i)} = -\frac{\ell_n(U) - \sum_{i=1}^n \log m(Y_i)}{n}.  \]
Also define $K^\star(U) = \inf\{K(m,m_{f,U}): f \in \FF\}$, where $\FF=\FF_U$ is the probability simplex in $\RR^{|U|}$.  This is the smallest Kullback--Leibler divergence of a mixture in the class $m_{f,U}$ from $m$.  Since $K(m,m_{n,U}) \to K^\star(U)$ for each $U$ and $K_n(U)$ is in some sense similar to $K(m,m_{n,U})$, one might expect that $K_n(U)$ also converges to $K^\star(U)$.  This will imply that maximizing the likelihood $\ell_n(U)$ in \eqref{eq:prml} to estimate $U$ is a reasonable strategy.  Indeed, \citet{pr-finite} proves that $K_n(U) \to K^\star(U)$ almost surely, as $n \to \infty$.  Furthermore, since the collection of all $U$'s is finite, the convergence is uniform, i.e., $K_n(\widehat U_n) \to K^\star(U^\star)$, where $\widehat U_n$ is the maximizer of $\ell_n(U)$ and $U^\star$ is the minimizer of $K^\star(U)$.  It follows that $\widehat U_n \to U^\star$ in the sense that, eventually, both sets will have the same elements.  Furthermore, Theorem~3 in \citet{pr-finite} implies that $f_{n,\widehat U_n}$ converges at a nearly $n^{-1/4}$ rate.  

Three remarks about this result are in order.  First, I had originally motivated SASA as a computationally efficient approximation to a Bayesian marginal likelihood procedure.  The convergence results above give SASA a secondary interpretation as a minimum distance method, not unlike those of \citet{james} and \citet{woosriram2006}.  Second, recall that \citet{chen1995} shows that the optimal rate for estimating finite mixing distributions with unknown support is $n^{-1/4}$, nearly matched by the rate for $f_{n,\widehat U_n}$.  However, I expect that this can be improved to $n^{-1/2}$, although the proof eludes me.  The difference comes from the fact that the essentially nonparametric problem of estimating the finite mixing distribution support is, here, first reduced to a very-high- but ultimately finite-dimensional parametric subproblem.  In fact, the rate $n^{-1/2}$ is available for the Bayesian marginal likelihood version.  Third, regarding estimation of the mixture complexity $|U|$, the results here differ considerably from those in, say, \citet{james} and \citet{woosriram2006}.  In particular, once the ``parameter space'' $\U$ is specified, the SASA estimate of the support size is bounded by $|\U|$, whether the model is correctly specified or not, and is guaranteed to converge.  In contrast, the \citet{james} and \citet{woosriram2006} estimates of the mixture complexity explode to infinity in the misspecified case.  I believe that, in the misspecified case, SASA's asymptotic identification of the best finite mixture in a sufficiently large class is more meaningful.  That is, one would arguably prefer the estimates to converge to the closest approximation of $m(y)$ within the postulated class of finite mixtures.

\subsection{Regularized SASA}
\label{SS:reg}

In the hierarchical model \eqref{eq:hierarchy}, it would be natural to introduce a prior for $U$ to complete the hierarchy.  \citet{mt-test} propose a regularized version of the the approximate marginal likelihood in which priors for structural parameters are incorporated into the model, effectively replacing the marginal likelihood with a marginal posterior.  

Here, a prior for $U$ should reflect the degree of sparsity in the mixture representation.  Since $S=|\U|$ will typically be large---much larger than the unknown support is likely to be---it is reasonable to penalize those $U$ with too many components.  To accomplish this, I recommend a prior for $U$ consisting of a binomial prior for the size $|U|$ and a conditionally uniform prior on $U$, given its size.  The parameters of the binomial prior are $(S,\rho)$, where $\rho$ denotes the prior probability that an element of $\U$ will be included in $U$.  The parameter $\rho$ can be adjusted to penalize candidate supports which are too large.  For example, one might be able to elicit an \emph{a priori} reasonable expected number of components, say 5, and then one may choose $\rho = 5/S$.  In the absence of such information, $\rho$ can be chosen by first estimating $m$ with some standard density estimate $\hat m$ and taking $\rho = (\text{number of modes of $\hat m$})/S$.  

\subsection{Numerical results}
\label{SS:results1}

For the simple univariate mixture problem, typical kernels are Gaussian with fixed scale and Poisson.  Example~\ref{ex:galaxy0} gives a quick comparison, in the context of Gaussian location mixtures, of SASA with the Bayesian method it is meant to approximate.  Example~\ref{ex:poismix.sim} gives the details of a large-scale simulation study for Poisson mixtures.  Location-scale mixtures (of Gaussians) will be the focus of the next section.  

\begin{ex}
\label{ex:galaxy0}
Consider the galaxy data set from \citet{roeder}.  This one consists of velocity measurements for $n=82$ galaxies.  Based on the \emph{a priori} considerations of \citet{escobar.west.1995}, it is reasonable to model these data as a location mixture of Gaussians with common scale $\sigma=1$.  The results for SASA, presented in Figure~1 of \citet{pr-finite}, using $\U=\{5.0,5.5,\ldots,39.5,40.0\}$, are obtained in roughly 3.5 seconds.  The fully Bayes version, using Jun Liu's sequential imputation algorithm to approximate the marginal likelihood $L_n^\dagger(U)$ in \eqref{eq:dp-marg}, gives identical results but takes more than 30 seconds.  Therefore, the approximate marginal likelihood $L_n(U)$ in \eqref{eq:prml} is reasonable.  
\end{ex}

\begin{ex}
\label{ex:poismix.sim}
Here I consider a simulation study presented in \citet{chen.khalili.2008} for Poisson mixtures.  There are seven models under consideration, and their respective parameters are listed in Table~\ref{table:pois.sim.pars}.  The resulting estimates of the mixture support size for a host of methods over 500 random samples each of size $n=100,500$ are summarized in Table~\ref{table:pois.sim}.  For the SASA implementation, I take $\Ubar = [0,20]$ and $\U$ a grid of $S=101$ equispaced points.  The SASA regularization parameter is taken as $\rho = 15 / S$.  For the most complicated Model~7, with $n=500$, the SASA estimate took about 15 seconds to compute, on average.  In addition to SASA, the methods compared are those based on AIC, BIC, and likelihood ratio test (LRT) model selection procedures, two minimum Hellinger distance (HD) methods from \cite{woosriram2007}, and the mixture SCAD method of \citet{chen.khalili.2008}.  The most striking observation is in Models 2--3 with $n=500$.  Note that existing methods estimate the support as a single point, while SASA is able to correctly identify two support points in a large proportion of the runs.  In Models 6--7, where the number of support points is relatively large, SASA tends to underestimate the support size.  But, in these examples, only MSCAD is successful, and SASA's performance better than BIC and the Woo--Sriram estimates.  
\end{ex}

\begin{table}
\begin{center}
{\small
\begin{tabular}{ccccc}
\hline
Model & $(u_1, f(u_1))$ & $(u_2, f(u_2))$ & $(u_3, f(u_3))$ & $(u_4, f(u_4))$ \\
\hline
1 & (1, .50) & (9, .50) & & \\
2 & (1, .80) & (9, .20) & & \\
3 & (1, .95) & (10, .05) & & \\
4 & (1, .45) & (5, .45) & (10, .10) & \\
5 & (1, .33) & (5, .33) & (10, .34) & \\
6 & (1, .30) & (5, .40) & (9, .25) & (15, .05) \\
7 & (1, .25) & (5, .25) & (10, .25) & (15, .25) \\
\hline
\end{tabular}
}
\end{center}
\caption{Parameters for the Poisson mixture simulations in Example~\ref{ex:poismix.sim}.}
\label{table:pois.sim.pars}
\end{table}

\begin{table}
\begin{center}
{\footnotesize
\begin{tabular}{cccccccccccccc}
& & & \multicolumn{5}{c}{$n=100$} & & \multicolumn{5}{c}{$n=500$} \\
\cline{4-8} \cline{10-14} 
Model & $|U|$ & Method & 1 & 2 & 3 & 4 & 5 & & 1 & 2 & 3 & 4 & 5 \\
\hline 
1 & 2 & AIC & & .938 & .062 & & & & & .942 & .076 & & \\
& & BIC & & .998 & .002 & & & & & .998 & .002 & & \\
& & HD$_{2/n}$ & & 1.00 & & & & & & 1.00 & & & \\
& & HD$_{\log n/n}$ & & 1.00 & & & & & & 1.00 & & & \\
& & LRT & & .950 & .050 & & & & & .960 & .040 & & \\
& & MSCAD & & .988 & .012 & & & & & 1.00 & & & \\
& & SASA & & .982 & .028 & & & & & .998 & .002 & & \\
\hline
2 & 2 & AIC & & .958 & .042 & & & & .950 & .042 & .008 & & \\
& & BIC & & .994 & .006 & & & & 1.00 & & & & \\
& & HD$_{2/n}$ & & .998 & .002 & & & & 1.00 & & & & \\
& & HD$_{\log n/n}$ & .002 & .998 & & & & & 1.00 & & & & \\
& & LRT & & .950 & .050 & & & & .960 & .040 & & & \\
& & MSCAD & .002 & .986 & .012 & & & & .990 & .010 & & & \\
& & SASA & & .988 & .012 & & & & & .972 & .028 & & \\
\hline
3 & 2 & AIC & .012 & .948 & .036 & & & & .950 & .048 & .002 & & \\
& & BIC & .026 & .972 & .002 & & & & .998 & .002 & & & \\
& & HD$_{2/n}$ & .616 & .384 & & & & & 1.00 & & & & \\
& & HD$_{\log n/n}$ & .946 & .054 & & & & & .994 & .006 & & & \\
& & LRT & & .930 & .070 & & & & .950 & .050 & & & \\
& & MSCAD & .052 & .868 & .080 & & & & .994 & .004 & & & \\
& & SASA & .024 & .974 & .002 & & & & & .932 & .068 & & \\
\hline
4 & 3 & AIC & & .410 & .590 & & & & & .006 & .972 & .022 & \\
& & BIC & & .778 & .222 & & & & & .100 & .900 & & \\
& & HD$_{2/n}$ & & .966 & .034 & & & & & .162 & .838 & & \\
& & HD$_{\log n/2}$ & & 1.00 & & & & & & .846 & .154 & & \\
& & LRT & & .390 & .580 & .020 & & & & .940 & .060 & & \\
& & MSCAD & & .280 & .692 & .028 & & & & .082 & .896 & .022 & \\
& & SASA & & .670 & .330 & & & & & .040 & .958 & .002 & \\
\hline
5 & 3 & AIC & & .274 & .720 & .006 & & & & & .974 & .026 & \\
& & BIC & & .684 & .316 & & & & & .026 & .974 & & \\
& & HD$_{2/n}$ & & .840 & .160 & & & & & .018 & .982 & & \\
& & HD$_{\log n/n}$ & & .988 & .012 & & & & & .462 & .538 & & \\
& & LRT & & .300 & .660 & .030 & & & & & .940 & .060 & \\
& & MSCAD & & .200 & .780 & .020 & & & & .016 & .964 & .020 & \\
& & SASA & & .436 & .554 & .010 & & & & .010 & .904 & .086 & \\
\hline
6 & 4 & AIC & & .080 & .878 & .042 & & & & & .644 & .356 & \\
& & BIC & & .316 & .680 & .004 & & & & .974 & .026 & & \\
& & HD$_{2/n}$ & & .718 & .282 & & & & & .956 & .044 & & \\
& & HD$_{\log n/n}$ & & .962 & .038 & & & & .060 & .940 & .538 & & \\
& & LRT & & .090 & .780 & .130 & & & & & .590 & .380 & .030 \\
& & MSCAD & & .010 & .666 & .320 & .004 & & & & .366 & .624 & .010 \\
& & SASA & & .194 & .804 & .002 & & & & & .892 & .108 &  \\
\hline
7 & 4 & AIC & & .010 & .918 & .072 & & & & & .592 & .408 & \\
& & BIC & & .134 & .858 & .008 & & & & .970 & .030 & & \\
& & HD$_{2/n}$ & & .182 & .812 & .006 & & & & .924 & .076 & & \\
& & HD$_{\log n/n}$ & & .718 & .282 & & & & .060 & 1.00 & & & \\
& & LRT & & .020 & .860 & .120 & & & & & .590 & .400 & .010 \\
& & MSCAD & & & .512 & .460 & .028 & & & & .110 & .812 & .078 \\
& & SASA & & .062 & .914 & .024 & & & & & .888 & .112 & \\
\hline
\end{tabular}
}
\end{center}
\caption{Results of the Poisson mixture simulations in Example~\ref{ex:poismix.sim}. Values are the proportion of estimates of the given size in 500 samples.  All but the SASA results are taken from \citet[][Tables~6--8]{chen.khalili.2008}}
\label{table:pois.sim}
\end{table}

\section{SASA for location-scale mixtures}
\label{S:locscale}

\subsection{Setup and modified algorithm}
\label{SS:approxPRML}

In principle, the SASA procedure is able to handle any type of finite mixture.  However, when $\U$ is a lattice in a higher-dimensional space, the computations become somewhat costly.  For a two-parameter kernel, for example, the approach outlined above would be to construct a lattice in the two-dimensional $u$-space and use the same in/out simulated annealing algorithm as in Section~\ref{SS:sann} for pairs $u=(u_1,u_2)$.  The collection $2^{\U}$ of all such pairs is, in general, quite large so it is advantageous to introduce a simpler approximation of the two-parameter mixture model.  My approach starts with the observation that, in general, the full two-parameter model could potentially have pairs $(u_1,u_2)$ and $(u_1,u_2')$ both entering the mixture.  The simplification is to rule out such cases, allowing at most one instance of, say, $u_1$ in the mixture.  This reduces the size of the search space and, in turn, accelerates the simulated annealing optimization step.  Here I develop this modification for the important special case of location-scale mixtures.  

Let $\Ubar_1$ and $\Ubar_2$ be closed intervals in $\RR$ and $\RR^+$, respectively, assumed to contain the range of values the location $\mu$ and scale $\sigma$ can take.  Chop up these intervals into sufficiently fine grids $\U_1$ and $\U_2$ of sizes $S_1 = |\U_1|$ and $S_2 = |\U_2|$, respectively.  Define the rectangle $\Ubar = \Ubar_1 \times \Ubar_2$ and the two-dimensional lattice $\U = \U_1 \times \U_2$.  Then the finite mixture model is just as before 
\[ m(y) = \sum_{(\mu,\sigma) \in U} p(y \mid \mu, \sigma) f(\mu,\sigma), \quad U \subset \U, \]
where the kernel $p(y \mid \mu, \sigma)$ equals $\sigma^{-1} p\bigl( (y-\mu)/\sigma \bigr)$ for some symmetric unimodal density function $p$.  This covers the case of finite location-scale Gaussian mixtures, but also the robust class of finite Student-t mixtures with a common fixed degrees of freedom.   Here I will focus on the Gaussian case only.

What makes this different from before is that $U$ can contain at most one $(\mu,\sigma)$ pair on each vertical line $\U_1 \times \U_2$.  To accommodate this restriction, I shall modify the simulated annealing algorithm proposed in Section~\ref{SS:sann}.  The key idea is to continue to use the location as the main parameter, but adjust the in/out scheme from before to allow for various levels of ``in.''  Recall the indicators $H_s$ in Section~\ref{SS:sann}.  Here I use the notation $H = (H_1,\ldots,H_{S_1})$, where each $H_s$ takes values in $\{0,1,\ldots,S_2\}$ to characterize the support set $U$.  The interpretation is 
\begin{equation}
\label{eq:Hvec}
H_s = \begin{cases} 0 & \text{if $\mu_s$ is not in the mixture} \\ h & \text{if pair $(\mu_s,\sigma_h)$ is in the mixture, $h=1,\ldots,S_2$.} \end{cases}
\end{equation}
In other words, location $\mu_s$ enters the mixture only if $H_s > 0$, but can enter paired with any of the scales $\sigma_h$ depending on the non-zero value of $H_s$.  Since there is a one-to-one correspondence between admissible subsets $U \subset \U$ and vectors $H$ of this form, I can formulate the SASA optimization problem in terms of $H$.  By restricting the estimates to this collection of admissible subsets, the state space to search goes from $2^{S_1 \times S_2}$ down to $(S_2 + 1)^{S_1}$, which can be a drastic reduction.  To maximize the approximate log marginal likelihood $\ell_n(H)$ over the set of all admissible $H$, I propose a modification of the foregoing simulated annealing algorithm.  In particular, the structure of the algorithm presented in Section~\ref{SS:sann} remains the same---all that changes is the proposal distribution.  

At iteration $t$, define $\beta(t) = S_1^{-1} \sum_{s=1}^{S_1} I\{H_s^{(t)} = 0\}$, the proportion of zero entries in $H^{(t)}$.  Now sample an entry in $H^{(t)}$ with probabilities
\begin{equation}
\label{eq:proposal}
\pi_s^{(t)} \propto 1 + \bigl( 1-\beta(t) \bigr)^{-r} \cdot I\{H_s^{(t)} > 0\}, \quad s=1,\ldots,S(\theta). 
\end{equation}
When $H^{(t)}$ has many zero entries, $1-\beta(t)$ will be small, so the non-zero entries will have greater chance of being sampled.  Let $H_s^{(t)}$ be the chosen entry.  To define $H_{\text{new}}$, there are two cases to consider:
\begin{itemize}
\item If $H_s^{(t)} = 0$, take $H_{\text{new}} \sim {\sf Unif}\{1,\ldots,S_2\}$.  
\vspace{-2mm}
\item If $0 < H_s^{(t)} < S_2$, take $H_{\text{new}} = 0$ with probability $\beta(t)$ and 
\[ H_{\text{new}} \sim {\sf Unif}\{H_s^{(t)}-1, H_s^{(t)} + 1\} \quad \text{with probability $1-\beta(t)$}. \]
If $H_s^{(t)} = 1$ or $S_2$, then $H_{\text{new}}$ would be 2 or $S_2-1$, respectively.  
\end{itemize}
The idea is to maintain the entry sampling that encourages a sparse mixture.  This is accomplished by, first, encouraging the selection of non-zero $H^{(t)}$ entries.  Second, these selected non-zero entries will likely be set to zero as the algorithm proceeds because $\beta(t)$ will tend to increase with $t$.  Thus, only the crucial components of $\U_1$ should remain in the mixture as $t$ increases.  

Once $H_{\text{new}}$ has been sampled, the simulated annealing algorithm decides to take $H^{(t+1)}$ as $H_{\text{new}}$ or $H^{(t)}$ depending on the flip of the $\alpha(t)$-coin as in Step 2 in Section~\ref{SS:sann}.  As before, if $H_{\text{new}}$ is a better candidate support than $H^{(t)}$ then the move will be accepted.  But allowing some moves to worse candidates helps prevent the simulated annealing procedure from getting stuck at local modes.

\subsection{Numerical results}
\label{SS:results3}

\begin{figure}
\begin{center}
\subfigure[Mixing distribution]{\scalebox{0.6}{\includegraphics{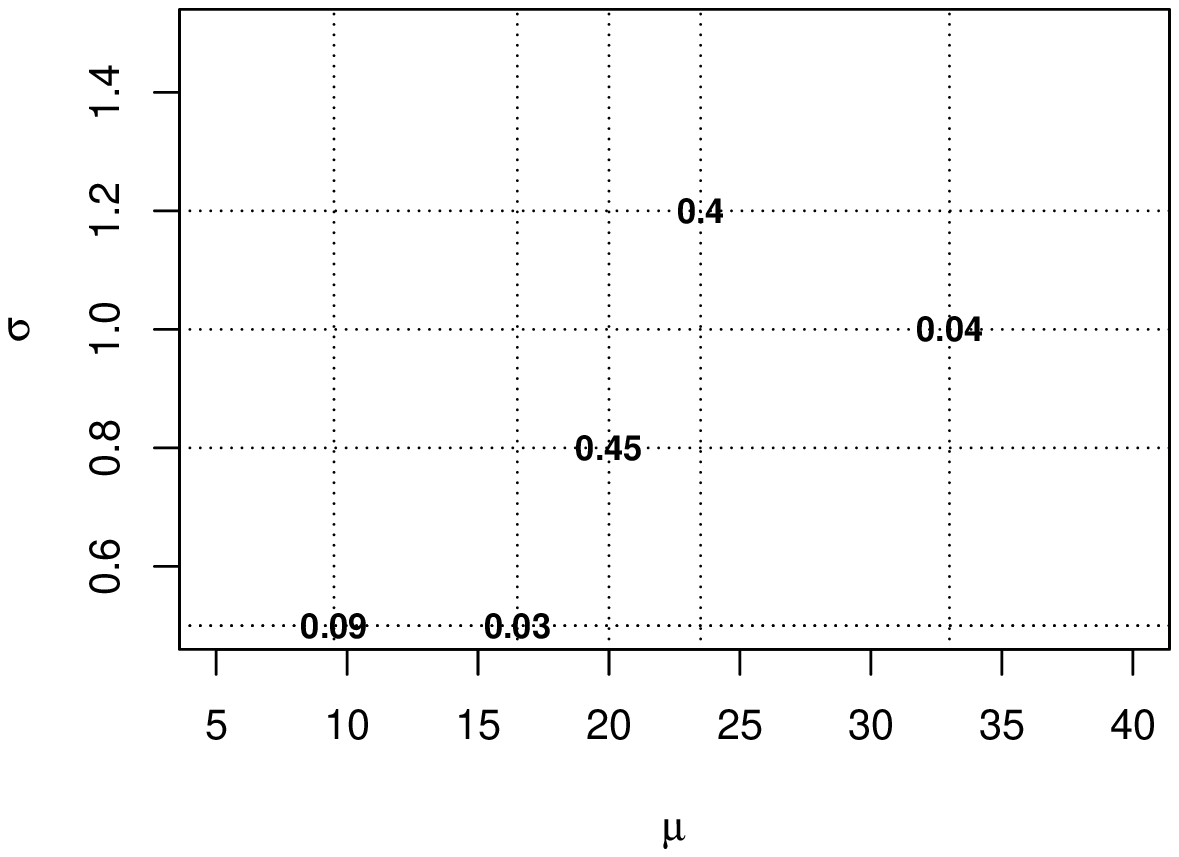}}}
\subfigure[Mixing distribution]{\scalebox{0.6}{\includegraphics{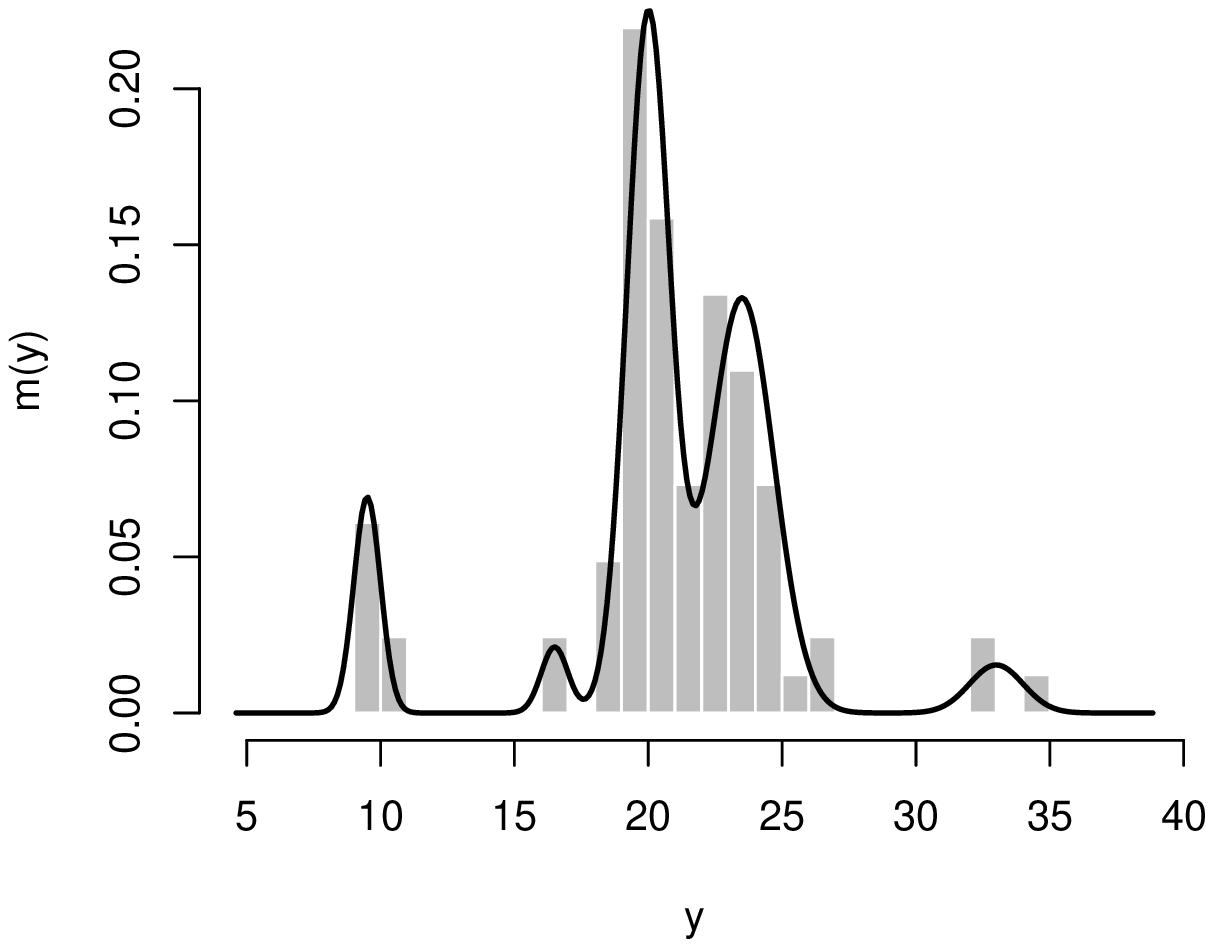}}}
\caption{Plot of the SASA estimates of the location-scale mixture density for the galactic velocity data in Example~\ref{ex:galaxy3}. In panel (a), the numbers are the mixing weights.}
\label{fig:ls.galaxy}
\end{center}
\end{figure}

\begin{ex}
\label{ex:galaxy3}
Here I use SASA to estimate a Gaussian location-scale mixture for the galaxy data in Example~\ref{ex:galaxy0}.  Take $\U_1=\{5.0,5.5,\ldots,39.5,40.0\}$ and $\U_2 = \{0.5,0.6,\ldots,1.4,1.5\}$.  SASA estimates five Gaussian components with varying scales, and Figure~\ref{fig:ls.galaxy} shows the resulting estimate of the density.  In this case, the overall fit is good---similar to that in \cite{ishwaran.james.sun.2001} and elsewhere---but only five components are needed compared to six in Example~1 in \citet{pr-finite}.  Here the computation took roughly 6 seconds. 
\end{ex}

\begin{ex}
\label{ex:gauss.sim2}
Next I present a simulation experiment in which the focus is on estimating the number of components in a challenging Gaussian mixture model considered in \citet{james} and \citet{woosriram2006}.  The particular model is 
\begin{equation}
\label{eq:weird.mixture}
m(y) = 0.25 {\sf N}(y \mid -0.3, 0.05) + 0.5 {\sf N}(y \mid 0, 10) + 0.25 {\sf N}(y \mid 0.3, 0.05). 
\end{equation}
The two components with variance $0.05$ makes for two nearby but dramatic modes.  With small sample sizes especially, it should be relatively difficult to detect these two distinct components.  For this model, accurate estimation of the number of components requires varying scale parameters, and I investigate the performance of the approximate SASA procedure outlined in Section~\ref{SS:approxPRML}.  

Table~\ref{table:sim2} summarizes the SASA estimates of the mixture complexity based on 100 random samples from the mixture model $m(y)$ in \eqref{eq:weird.mixture} with four different sample sizes: $n = 50$, 250, 500, and 1000.  In particular, I take $\Ubar_1 = [-2,2]$, $\Ubar_2 = [0.1, 4.0]$ and $\U_1$ and $\U_2$ are equispaced grids of length $S_1 = 40$ and $S_2 = 25$, respectively.  Note that the true location and scale parameters in \eqref{eq:weird.mixture} do not belong to $\U_1 \times \U_2$.  The simulated annealing optimization procedure in Section~\ref{SS:approxPRML} is used to optimize the approximate marginal likelihood over the collection of admissible subsets, which provides an estimate of the mixture complexity.  In this case, there are $2^{40 \times 25} \approx 10^{301}$ subsets of $\U_1 \times \U_2$, compared to $26^{40} \approx 4 \times 10^{56}$ admissible subsets, so there is a substantial computational savings in using the approximation in Section~\ref{SS:approxPRML}.  The average computation time for SASA ranges from 4.5 seconds for $n=50$ and 52 seconds for $n=1000$.  For comparison, I also include minimum distance estimates of \citet{woosriram2006} and \citet{james}, and the Bayesian estimates of \citet{roederwasserman}; these shall be denoted by \emph{WS}, \emph{JPM}, and \emph{RW}, respectively.  The RW method performs well for small $n$ but seems to falter as $n$ increases, while the JPM method does well for large $n$.  SASA does quite well for $n=50$ and, although it is not the best, it is competitive in all other cases.  In particular, it seems that only the WS method is as good or better than SASA at correctly identifying the true mixture complexity across simulations.  
\end{ex}

\begin{table}
\begin{center}
{\small
\begin{tabular}{llrrrrrrrr}
 & & \multicolumn{8}{c}{Number of components} \\
\cline{3-10}
 & Method & 1 & 2 & 3 & 4 & 5 & 6 & 7 & 8 \\
\hline
$n=50$ & WS & 80 & 20 & & & & & & \\
& JPM & 44 & 53 & 3 & & & & & \\
& RW & 22 & 7 & 59 & 10 & 1 & 1 & & \\
& SASA & & 23 & 44 & 25 & 8 & & & \\
\hline
$n=250$ & WS & 16 & 39 & 45 & & & & & \\
& JPM & & 87 & 11 & 1 & 1 & & & \\
& RW & & & 60 & 22 & 18 & & & \\
& SASA & 15 & 28 & 47 & 17 & 1 & & & \\
\hline
$n=500$ & WS & & 35 & 65 & & & & & \\
& JPM & & 58 & 34 & 6 & 2 & & & \\
& RW & & & 22 & 12 & 61 & 5 & & \\
& SASA & & 17 & 48 & 32 & 2 & 1 & & \\
\hline
$n=1000$ & WS & & 26 & 74 & & & & & \\
& JPM & & 18 & 63 & 10 & 2 & 3 & 1 & 3 \\
& RW & & & & 1 & 89 & 10 & & \\
& SASA & & 10 & 50 & 24 & 13 & 3 & & \\
\hline
\end{tabular}
}
\end{center}
\caption{Summary of the 100 estimates of $|U|$ in Example~\ref{ex:gauss.sim2}. The true mixture complexity is 3.  All but the SASA results are taken from \citet[][Table~1]{woosriram2006}.}
\label{table:sim2}
\end{table}

\section{Discussion}
\label{S:discuss}

This paper presents a novel hybrid stochastic approximation--simulated annealing algorithm for estimating finite mixtures.  The method is based, first, on a marginal likelihood function for the support based on a Bayesian hierarchical model.  Then two approximations are introduced to estimate the unknown support $U$---the first is an approximation of the bounding set $\Ubar \supseteq U$ by a finite grid $\U$, and the second is an efficient approximation of the marginal likelihood.  Then a simulated annealing algorithm is used to maximize this approximate marginal likelihood over the finite set of candidate $U$'s.  There may be some theoretical benefits, in terms of rates of convergence, to approximating $\Ubar$ with the finite set $\U$, but the details remain to be worked out.  Examples in both the Poisson and Gaussian case indicate that SASA is competitive with existing methods.  In my experience, SASA is generally a bit more expensive computationally than the minimum distance methods of \citet{woosriram2006} or \citet{james}.  But, on the other hand, it tends to be faster than the Bayesian methods it is meant to approximate.  

In applications, the initial choice of $\U$ and, in particular, $|\U|$ is not obvious.  In practice, one should make this choice based on the shape/spread of the data and the chosen kernel; this was the approach taken in Examples~\ref{ex:galaxy0} and \ref{ex:galaxy3}.  An interesting proposition is to let $\U = \U_n$ depend on the sample size $n$, like a sieve.  The idea is that, if $\U$ is sufficiently large, then the class of mixtures supported on subsets of $\U$ should be rich enough to closely approximate $m$.  For example, suppose that $m$ is a finite mixture with support points somewhere in the compact bounding set $\Ubar$.  Then it should be possible to choose $\U_n$ to saturate the bounding set $\Ubar$ at a suitable rate so that $K(m,m_{n,\widehat U_n}) \to 0$ almost surely.  To prove this, bounds on the constants associated with the rate in \citet{pr-finite} would be needed, since these would likely depend on $|\U|$.  

An explicit penalty on the size of the mixing distribution support was introduced in Section~\ref{SS:reg}.  And the location-scale adjustment to SASA's simulated annealing proposal can also be viewed as an implicit penalty on $U$.  An anonymous reviewer pointed out the potential for incorporating even more sophisticated penalties in the approximate marginal likelihood for $U$.  For example, one could further penalize candidate supports that contain points which are too close in some sense.  This extreme regularization was not necessary in the examples considered here, but if the grid $\U$ is very fine, then the the closeness of nearby support points may become a concern.  

In some cases, one might want to consider, say, a Gaussian location mixture with fixed but unknown scale $\sigma$.  It is straightforward to implement an intermediate step in the algorithm in Section~\ref{SS:sann} whereby one replaces the joint marginal likelihood $\ell_n(U,\sigma)$ by a profile version $\ell_n(U,\hat\sigma)$.  In my experience, this was actually more expensive computationally than the location-scale approach, so I did not pursue this direction. 


\section*{Acknowledgments}

The author is grateful to two anonymous reviewers for their thoughtful criticisms that led to an overall improvement of the paper.  Also special thanks go to Professors J.~K.~Ghosh, Surya T.~Tokdar, and Chuanhai Liu for a number of helpful suggestions.  A portion of this work was completed while the author was affiliated with the Department of Mathematical Sciences, Indiana University--Purdue University Indianapolis.

\bibliographystyle{/Users/rgmartin/Research/TexStuff/asa}
\bibliography{/Users/rgmartin/Research/mybib}

\end{document}